\documentclass[runningheads]{llncs}
\usepackage{graphicx}
\usepackage{booktabs} % For formal tables
\usepackage{amssymb}
\usepackage{multirow}
\usepackage{amsmath}
\usepackage{paralist}
\usepackage{tikz}
\usepackage{colortbl}
\usepackage{todonotes}
\usepackage{color}
\begin{document}
\title{DeepTagRec: A Content-cum-User based\\ Tag Recommendation Framework for Stack Overflow}
\titlerunning{DeepTagRec: A Tag Recommendation Framework}

\author{Suman Kalyan Maity$^1$\thanks{Most of the work was done when all the authors were at IIT Kharagpur, India. We also acknowledge Prithwish Mukherjee, Shubham Saxena, Robin Singh, Chandra Bhanu Jha for helping us in various stages of this project.}, Abhishek Panigrahi$^2$, Sayan Ghosh$^2$, Arundhati Banerjee$^2$, Pawan Goyal$^2$ and Animesh Mukherjee$^2$}
\institute{$^1$Northwestern University; $^2$Dept. of CSE, IIT Kharagpur, India \email{suman.maity@kellogg.northwestern.edu}}

\authorrunning{Maity et al.}

%
\maketitle              % typeset the header of the contribution

\begin{abstract}
In this paper, we develop a {\em content-cum-user} based deep learning framework $DeepTagRec$ to recommend appropriate question tags on Stack Overflow. The proposed system learns the content representation from question title and body. Subsequently, the learnt representation from heterogeneous relationship between user and tags is fused with the content representation for the final tag prediction. On a very large-scale dataset comprising half a million question posts, $DeepTagRec$ beats all the baselines; in particular, it significantly outperforms the best performing baseline $TagCombine$ achieving an overall gain of 60.8\% and 36.8\% in $precision@3$ and $recall@10$ respectively. $DeepTagRec$ also achieves 63\% and 33.14\% maximum improvement in \textit{exact-k accuracy} and \textit{top-k accuracy} respectively over $TagCombine$.

\end{abstract}
\vspace{-4mm}
\section{Introduction}
In community based question answering (CQA) websites like Yahoo! Answers, Stack Overflow, Ask.com, Quora etc., users generate content in the form of questions and answers, facilitating the knowledge gathering through collaboration and contributions in the Q\&A community. These questions are annotated with a set of tags by users in order to topically organize them across various subject areas. The tags are a form of metadata for the questions that help in indexing, categorization, and search for particular content based on a few keywords. Hashtags in social media or CQA tags are precursor to folksonomy or social/collaborative tagging (Del.icio.us\footnote{\url{http://del.icio.us}}, Flickr\footnote{\url{http://www.flickr.com}}). The tagging mechanism in folksonomy is fully crowd-sourced and unsupervised and hence the annotation and the overall organization of tags suffers from uncontrolled use of vocabulary resulting in wide variety of tags that can be redundant, ambiguous or entirely idiosyncratic. Tag ambiguity arises when users apply the same tag in different contexts which gives the false impression that resources are similar when they are in fact unrelated. Tag redundancy, on the other hand, arises when several tags bearing the same meaning are used for the same concept. Redundant tags can hinder algorithms that depend on identifying similarities between resources. Further, manual error and malicious intent of users could also lead to improperly tagged questions, thereby, jeopardizing the whole topical organization of the website. A tag recommendation system can help the users with a set of tags from where they can choose tags which they feel best describe the question, thus facilitating faster annotations. Moreover, tag recommendation decreases the possibility of introducing synonymous tags into the tag list due to human error, thereby, reducing tag redundancy in the system.\\
The problem of tag recommendation has been studied by various researchers from various different perspectives~\cite{Heymann,Fu:2008,Song,lu,Krestel,Rendle,Liu:2011,liu2012,ding,Godin,gong}.~\cite{lu} proposes a content-based method that incorporates the idea of tag/term coverage while~\cite{Song} proposes a two-way Poisson mixture model for real-time prediction of tags.~\cite{sen} proposes a user-based vocabulary evolution model.~\cite{hu2011} present a framework of personalized tag recommendation in Flickr using social contacts.~\cite{Krestel} propose a LDA based method for extracting a shared topical structure from the collaborative tagging effort of multiple users for recommending tags.~\cite{Rendle} presents a factorization model for efficient tag recommendation.~\cite{Liu:2011} build a word trigger method to recommend tags and further use this framework for keyphrase extraction~\cite{liu2012}.~\cite{nie} leverage similar questions to suggest
tags for new questions. A recent paper by Wu et al.~\cite{wu} exploits question similarity, tag similarity and tag importance and learns them in a supervised random walk framework for tag recommendation in Quora. Further, Joulin et al.~\cite{joulin2016bag} proposes a general text classification which we have adapted for comparison with our model.\if{0}~\cite{ding} propose a topical translation model for hashtag suggestion.\fi~\cite{Godin} develops an unsupervised content-based hashtag recommendation for tweets while~\cite{She} proposes a supervised topic model based on LDA.~\cite{gong} uses Dirichlet process mixture model for hashtag recommendation.~\cite{Ma} proposes a PLSA-based topic model for hashtag recommendation. Weston et al.~\cite{tagspace} proposes a CNN-based model for hashtag prediction.\\
In this paper, we employ a {\em content-cum-user} based deep learning framework for tag recommendation model which takes advantage of the content of the question text and is further enhanced by the rich relationships among the users and tags in Stack Overflow. We compare our method with existing state-of-the-art methods like Xia et al.~\cite{wang15}, Krestel et al.~\cite{Krestel}, Wu et al.~\cite{wu}, Lu et al.~\cite{lu}, Joulin et al.'s \textit{fastText}~\cite{joulin2016bag} and Weston's \textit{\#TAGSPACE} method~\cite{tagspace} and observe that our method performs manifold better.   

\section{Tag Recommendation Framework}
\begin{figure}
\centering
\includegraphics[scale = 0.35]{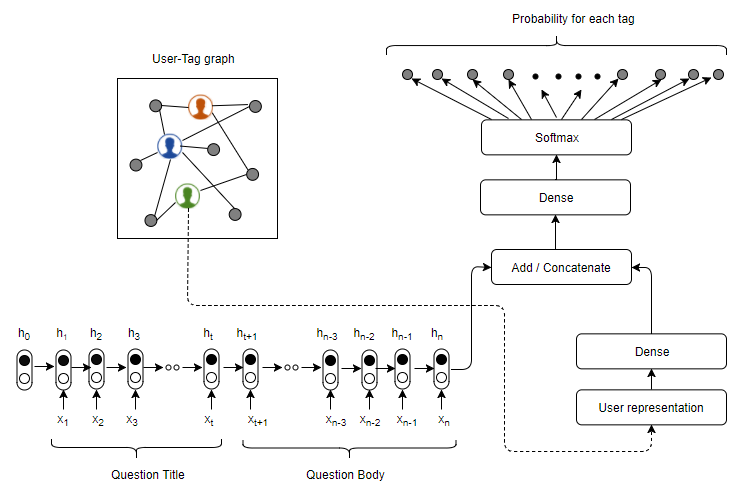}
\caption{$DeepTagRec$ tag recommendation framework }
\label{framework_diag}
\end{figure}

In this section, we describe in detail the working principles of our proposed recommendation system $DeepTagRec$\footnote{The codes and data are available at https://bit.ly/2HsVhWC}. The basic architecture of the model is shown in Fig.~\ref{framework_diag}. The whole framework consists of three major components: (a) content representation extraction from question title and body, (b) learning user representation from heterogeneous user-tag network using node2vec~\cite{node2vec-kdd2016}, (c) tag prediction using representation aggregation. We formulate the tag prediction model as a function of the content and the user information. 
%% give fig number
\vspace{-10mm}
\subsection{Content representation}
To obtain the representation of body and title, we use Gated Recurrent Unit (GRU) model to encode the content as a sequence of words. Given the title of the question $T$ and main body of the question $B$, we first run a GRU to learn representation of $T$, denoted by $c_T$. In the next step, we learn the representation of $B$ denoted by $c_B$ using a GRU, having $c_T$ as the initial hidden state. We then describe the inner mechanism of a GRU. The GRU hidden vector output at step $t$, $h_t$, for the input sequence $X$ = $(x_1, ...,x_t, ...,x_n)$ is given by:
\begin{equation}
\begin{aligned}
&z_t = \sigma(W_z x_t + U_z h_{t-1}) \\
&r_t = \sigma(W_r x_t + U_r h_{t-1}) \\
&\tilde{h_t} = tanh(W_h x_t + U_h (r_t \odot h_{t-1})) \\
&h_t = z_t \odot \tilde{h_t} + (1 - z_t) \odot h_{t-1} 
\end{aligned}
\end{equation}

where, $W_z$, $W_r$, $W_h$ $\in$ $R^{m \times d}$ and $U_z$, $U_r$, $U_h$ $\in$ $R^{d \times d}$ are the weight vectors, $m$ and $d$ denote the word2vec dimension of word $x_t$ and hidden state $h_t$ respectively, $z_t$ and $r_t$ denote the update gate and reset gate in the GRU. The initial state $h_{0}$ is either vector 0 or is given as an input. We shall denote the entire GRU model by G(X, $h_{0}$) for future references.

Let $T$ and $B$ denote the sequence of words in the title and body of the question, respectively. Each word present in the sequence has its word2vec~\cite{mikolov2013efficient} representation. In case the predefined word vector does not exist, we consider a 300 dimension zero vector for the word. So, the content representation can be summarized as $c_T = G(T, 0); c_B = G(B, c_T)$

\subsection{User-tag network representation}
We construct a heterogeneous network containing nodes corresponding to users and tags. Let the graph be denoted by $G(V, E)$ where $V = V_{U} \cup V_{T}$, $V_{U}$ are nodes corresponding to users and $V_{T}$ are nodes corresponding to tags. We add an edge between a user and a tag, if the user has posted some question with the tag present in the question's tagset. The basic idea is to create a network and understand the tag usage pattern of each user. Given this graph $G(V, E)$, we use node2vec (a semi-supervised algorithm for learning feature representation of nodes in the network using random-walk based neighborhood search) to learn the representation of each node present in the graph. Let
$f \colon V \rightarrow$ $R^d$
be the mapping function from nodes to feature representations. Node2vec optimizes the following objective function, which maximizes the log-probability of observing a network neighborhood $N_S(u)$ for a node $u$
conditioned on its feature representation, 
$$
\max_{f} \sum_{u \in V} log (P(N_S(u) | u)) 
$$
where $P(N_S(u) | u)$ is given by
$$
P(N_S(u) | u) = \prod_{n_i \in N_S(u)} \frac{\exp(f(n_i) f(n_u))}{\sum_{v \in V} \exp(f(v) f(n_u))} 
$$

Node2vec starts with a random function that maps each node to an embedding in $R^d$. This function is refined in an iterative way, so that the conditional probability of occurrence of the neighborhood of a node increases. The conditional probability of a node in the neighborhood of another node is proportional to cosine similarity of their embeddings. This is the idea that we use in our model, a user's representation should have high similarity with his/her adjacent tag nodes in the graph. 

\subsection{Representation aggregation and tag prediction}
Once we obtain both the word2vec representation ($Q_w$) of the question data and the node2vec representation ($U_n$) of users, we can aggregate these embeddings into final heterogeneous embedding ($f_{agg}$) by specific aggregation function $g(., .)$ as follows.
\begin{itemize}
\item Addition : $f_{agg} = Q_w + U_n$
\item Concatenation : $f_{agg} = [Q_w, U_n]$
\end{itemize}
Following this layer, we have a dense or fully connected layer and finally a sigmoid activation is applied in order to get a probability distribution over the 38,196 tags.

\vspace{-3mm}
\section{Evaluation}
In this section, we discuss in detail the performance of $DeepTagRec$ and compare it with six other recent and popular recommendation systems -- (i) Xia et al.'s $TagCombine$~\cite{wang15}, (ii) Krestel et al.'s~\cite{Krestel} LDA based approach, (iii) Lu et al.'s~\cite{lu} content based approach, (iv) Wu et al.'s~\cite{wu} question-tag similarity driven approach, (v) Weston et al.'s CNN-based model (\#TAGSPACE) and (vi) Joulin et al.'s fastText model~\cite{joulin2016bag}.\\
\textbf{Training and testing:}
We have 0.5 million training questions. Each question has different number of tags associated with it. 
Maximum length of the question is fixed as 300 words\footnote{Avg. length of questions is 129 words. For question length $<$300, we pad them with zero vectors.}. Each word is represented as a 300(m) dimension vector by using the predefined word2vec embeddings. The tags are represented as one-hot vectors. For a training example with \textit{t} tags we add these \textit{t} one hot vectors as the output for that training example. The number of GRU units is taken as 1000(d). The learning rate is taken to be 0.001 and the dropout as 0.5. For testing the model, we use 10K questions and perform the same initial steps of concatenating the title and body and then representing the words in 300 dimension vector form. For learning user representation, we create a user-tag graph over the training examples. We use node2vec over this graph to learn a 128 vector user embedding for all the users present in the training dataset. The output of the joint model is a probability distribution over 38,196 tags. We take the $k$ tags with the highest probability for further evaluation.
\vspace{-3mm}
\subsection*{Experiments and results}
In this section, we discuss in detail the performance of our proposed model $DeepTagRec$ and compare it against the baselines. \iffalse six different baselines $TagCombine$, \textit{Wu et al.}, \textit{Lu et al.}, \textit{Krestal et al.}, \textit{\#TAGSPACE} and \textit{fastText}.\fi To understand the effect of content and user information separately, we also experiment with a variant of the proposed model -- $DeepTagRec_{content}$ (i.e, only the content representation module of $DeepTagRec$). \iffalse For evaluation purpose, we use the following metrics - i) $precision@k$ and ii) $recall@k$.\fi
For evaluation purpose, we have used the following metrics.\\
\textbf{$Precision@k$} : Suppose there are $q$ questions and for each question $i$, let $TagU_i$ be the set of tags given by the asker to the question and $TagR_i$ be the set of top-$k$ ranked tags recommended by the algorithm, then $Precision@k = \frac{1}{q}\sum_{i=1}^{q}\frac{|TagU_i \cap TagR_i|}{|TagR_i|}$.
% \begin{equation*}
%  Precision@k = \frac{1}{q}\sum_{i=1}^{q}\frac{|TagU_i \cap TagR_i|}{|TagR_i|}
% \end{equation*}

\textbf{$Recall@k$} : Similarly as $precision@k$, it can be formally defined as $Recall@k = \frac{1}{q}\sum_{i=1}^{q}\frac{|TagU_i \cap TagR_i|}{|TagU_i|}$ where $k$ is a tunable parameter that determines how many tags the system recommends for each question.\\
$DeepTagRec$ significantly outperforms all the baselines (see Table~\ref{table:precrec_2}) obtaining a $precision@3$ of $\sim0.51$ and a $recall@10$ of $\sim 0.76$. Comparing the proposed variants, we observe that while most of the improvement of our model comes from the content representation, user information consistently helps in improving the performance. Since \textit{Wu et al.}, \textit{Lu et al.}, \textit{Krestal et al.}, \textit{\#TAGSPACE} and \textit{fastText} methods perform significantly worse, we have not considered them for subsequent analysis in this paper. We also do not consider $DeepTagRec_{content}$ further, and only compare $DeepTagRec$ with the best performing baseline, $TagCombine$.
\begin{table}[h]
\vspace{-1mm}
\centering
\caption{Precision (P) and Recall (R) in \iffalse Phase-II evaluation \fi for $DeepTagRec$ and the other baselines.}
\label{table:precrec_2}
\vspace{-1mm}
\resizebox{8.5cm}{!}{
\begin{tabular}{|c|c|c|c||c|c|c|}
\hline
Model & P@3 & P@5 & P@10& R@3 & R@5 & R@10 \\ \hline
% \textit{Krestel et al.}& 0.0862& 0.0722 & 0.0575 & 0.0906 & 0.1284 & 0.2075 \\ \hline
\textit{Krestel et al.} [2009]& 0.0707& 0.0603 & 0.0476 & 0.0766 & 0.1097& 0.1738\\ \hline
\textit{Lu et al.}[2009] & 0.1767 & 0.1351 & 0.0922&0.1952& 0.2477& 0.3362 \\ \hline
  \textit{Wu et al.} [2016] & 0.21 & 0.16&0.106&0.2325&0.2962&0.3788 \\\hline
  \textit{\#TAGSPACE} [2014] &0.105 & 0.087 & 0.063 & 0.111 & 0.162 & 0.511 \\\hline
  \textit{fastText} [2016] & 0.102 & 0.0783 & 0.149 & 0.0388 & 0.149 & 0.227 \\\hline
%\textit{Nguyen et al.} [2017] & 3.33e-05 & 2e-5 & 8e-5 & 5e-5 & 5e-5 & 23.17e-5 \\\hline 
\rowcolor{red!20}$TagCombine$ & 0.3194  & 0.2422  & 0.1535 & 0.3587 & 0.4460  & 0.5565   \\ \hline
% \rowcolor{green!20}$DeepTagRec_{content}$ & 0.4768&0.3457&0.2015&0.5388&0.6356&0.7270\\ \hline
% % $TagM$ & {\bf 0.3319}  & 0.2539  & 0.1651 & 0.3725 & 0.4666  & {\bf 0.5954}   \\ \hline
% \rowcolor{green}$DeepTagRec$ & \textbf{0.4948}&\textbf{0.3602}&\textbf{0.2099}&\textbf{0.5583}&\textbf{0.6594}&\textbf{0.7526}  \\ \hline
\rowcolor{green!20}$DeepTagRec_{content}$ & 0.4442 & 0.3183 & 0.184 & 0.5076 & 0.591 & 0.6702\\ \hline
% $TagM$ & {\bf 0.3319}  & 0.2539  & 0.1651 & 0.3725 & 0.4666  & {\bf 0.5954}   \\ \hline
\rowcolor{green}$DeepTagRec$ & \textbf{0.5135} & \textbf{0.3684}& \textbf{0.2125}& \textbf{0.5792}& \textbf{0.6736}& \textbf{0.7613}  \\ \hline
\end{tabular}}
\vspace{-1mm}
\end{table}

\noindent\textit{\textbf{Top-k and exact-k accuracy:}} Apart from precision and recall, we also define the following evaluation metrics for further comparison.
\begin{description}
\item [] \textit{Top-k accuracy} : This metric is defined as the fraction of questions correctly annotated by at least one of the $top-k$ tags recommended by the algorithm.
\item [] \textit{Exact-k accuracy}: This metric is defined as the fraction of questions correctly annotated by the $k^{th}$ recommended tag.
\end{description}
Table~\ref{table:topk} shows the top-$k$ and exact-$k$ accuracy for both the models and we can observe that $DeepTagRec$ outperforms $TagCombine$ by 33.14\%, 22.89\% and 13.5\% for $k$ = 3, 5 and 10 respectively w.r.t top-$k$ accuracy. $DeepTagRec$ also performs better in exact-$k$ accuracy than $TagCombine$ by achieving maximum and minimum gains of 63\% and 10\% respectively.
\begin{table}[h!]
\vspace{-1mm}
\centering
\caption{Top-$k$ and exact-$k$ accuracy. Values at first 3 columns are for top-$k$ accuracy and rest are for exact-$k$ accuracy.}
\label{table:topk}\small
\vspace{-3mm}
\resizebox{8.5cm}{!}{
\begin{tabular}{|c|c|c|c|||c|c|c|c|c|}
\hline
Model & $k = 3$ & $k = 5$ & $k = 10$ & $k = 1$ & $k = 2$ & $k = 3$ & $k = 4$ & $k = 5$\\ \hline
\rowcolor{red!20}$TagCombine$ & 0.688 &0.769&0.851 & 0.481&0.289&0.188 & 0.145 & 0.108 \\ \hline
% \rowcolor{green}$DeepTagRec$ & \textbf{0.893}&\textbf{0.932}&\textbf{0.961}&\textbf{0.713}&\textbf{0.469}&\textbf{0.302}&\textbf{0.190}&\textbf{0.127} \\ \hline
\rowcolor{green}$DeepTagRec$ & \textbf{0.916} & \textbf{0.945} & \textbf{0.966} & \textbf{0.784} & \textbf{0.468} & \textbf{0.289} & \textbf{0.184}& \textbf{0.118} \\ \hline
\end{tabular}}
\vspace{-1mm}
\end{table}

\vspace{-3mm}
\section{Conclusions}
In this paper, we propose a neural network based model ($DeepTagRec$) that leverages both the textual content (i.e., title and body) of the questions and the user-tag network for recommending tag. Our model outperforms the most competitive baseline $TagCombine$ significantly. We improve -- $precision@3$ by 60.8\%, $precision@5$ by 52.1\%, $precision@10$ by 38.4\%, $recall@3$ by 61.5\%, $recall@5$ by 51.03\%, $recall@10$ by 36.8\% -- over $TagCombine$. $DeepTagRec$ also performs better in terms of other metrices where it achieves 63\% and 33.14\% overall improvement in \textit{exact-k accuracy} and \textit{top-k accuracy} respectively over $TagCombine$. 

\bibliographystyle{splncs04}
\bibliography{ref}

\begin{thebibliography}{10}
\providecommand{\url}[1]{\texttt{#1}}
\providecommand{\urlprefix}{URL }
\providecommand{\doi}[1]{https://doi.org/#1}

\bibitem{ding}
Ding, Z., Qiu, X., Zhang, Q., Huang, X.: Learning topical translation model for
  microblog hashtag suggestion. In: IJCAI (2013)

\bibitem{Fu:2008}
Fu, W.T.: The microstructures of social tagging: A rational model. In: CSCW.
  pp. 229--238 (2008)

\bibitem{Godin}
Godin, F., Slavkovikj, V., De~Neve, W., Schrauwen, B., Van~de Walle, R.: Using
  topic models for twitter hashtag recommendation. pp. 593--596 (2013)

\bibitem{gong}
Gong, Y., Zhang, Q., Huang, X.: Hashtag recommendation using dirichlet process
  mixture models incorporating types of hashtags pp. 401--410 (2015)

\bibitem{node2vec-kdd2016}
Grover, A., Leskovec, J.: node2vec: Scalable feature learning for networks. In:
  KDD (2016)

\bibitem{Heymann}
Heymann, P., Ramage, D., Garcia-Molina, H.: Social tag prediction. In: SIGIR.
  pp. 531--538 (2008)

\bibitem{hu2011}
Hu, J., Wang, B., Tao, Z.: Personalized tag recommendation using social
  contacts. In: Proc. of SRS '11, in conjunction with CSCW. pp. 33--40 (2011)

\bibitem{joulin2016bag}
Joulin, A., Grave, E., Bojanowski, P., Mikolov, T.: Bag of tricks for efficient
  text classification. arXiv preprint arXiv:1607.01759  (2016)

\bibitem{Krestel}
Krestel, R., Fankhauser, P., Nejdl, W.: Latent dirichlet allocation for tag
  recommendation. In: RecSys. pp. 61--68 (2009)

\bibitem{Liu:2011}
Liu, Z., Chen, X., Sun, M.: A simple word trigger method for social tag
  suggestion. In: EMNLP. pp. 1577--1588 (2011)

\bibitem{liu2012}
Liu, Z., Liang, C., Sun, M.: Topical word trigger model for keyphrase
  extraction. In: COLING. pp. 1715--1730 (2012)

\bibitem{lu}
Lu, Y.T., Yu, S.I., Chang, T.C., Hsu, J.Y.j.: A content-based method to enhance
  tag recommendation. In: IJCAI. vol.~9, pp. 2064--2069 (2009)

\bibitem{Ma}
Ma, Z., Sun, A., Yuan, Q., Cong, G.: Tagging your tweets: A probabilistic
  modeling of hashtag annotation in twitter. In: CIKM. pp. 999--1008 (2014)

\bibitem{mikolov2013efficient}
Mikolov, T., Chen, K., Corrado, G., Dean, J.: Efficient estimation of word
  representations in vector space. arXiv preprint arXiv:1301.3781  (2013)

\bibitem{nie}
Nie, L., Zhao, Y.L., Wang, X., Shen, J., Chua, T.S.: Learning to recommend
  descriptive tags for questions in social forums. ACM TOIS  \textbf{32}(1), ~5
  (2014)

\bibitem{Rendle}
Rendle, S., Schmidt-Thieme, L.: Pairwise interaction tensor factorization for
  personalized tag recommendation. In: WSDM. pp. 81--90 (2010)

\bibitem{sen}
Sen, S., Lam, S.K., Rashid, A.M., Cosley, D., Frankowski, D., Osterhouse, J.,
  Harper, F.M., Riedl, J.: Tagging, communities, vocabulary, evolution. pp.
  181--190 (2006)

\bibitem{She}
She, J., Chen, L.: Tomoha: Topic model-based hashtag recommendation on twitter.
  In: WWW Companion. pp. 371--372 (2014)

\bibitem{Song}
Song, Y., Zhuang, Z., Li, H., Zhao, Q., Li, J., Lee, W.C., Giles, C.L.:
  Real-time automatic tag recommendation. In: SIGIR. pp. 515--522 (2008)

\bibitem{wang15}
Wang, X.Y., Xia, X., Lo, D.: Tagcombine: Recommending tags to contents in
  software information sites. J. Comput Sci.  \textbf{30}(5),  1017--1035
  (2015)

\bibitem{tagspace}
Weston, J., Chopra, S., Adams, K.: {\#}tagspace: Semantic embeddings from
  hashtags. In: EMNLP. pp. 1822--1827 (2014)

\bibitem{wu}
Wu, Y., Wu, W., Li, Z., Zhou, M.: Improving recommendation of tail tags for
  questions in community question answering. In: AAAI (2016)

\end{thebibliography}

\end{document}